\documentclass[doublecol,figures]{epl2}
\usepackage{amssymb}
\usepackage{graphicx}
\usepackage{amsmath}

\shorttitle{}
\institute{
  \inst{1} Department of Physics, Harvard University, Cambridge, MA 02138-2933 USA\\
  \inst{2} Department of Physics, Virginia Tech, Blacksburg, VA 24061-0435 USA
}
\abstract{We consider a system in a non-equilibrium steady state by joining
two semi-infinite Ising chains coupled to thermal reservoirs with {\em
different} temperatures, $T$ and $T^{\prime}$.  To compute the energy flux from the hot bath through our system into the cold bath, we exploit Glauber heat-bath dynamics to derive an exact equation for the two-spin correlations, which we solve for $T^{\prime }=\infty$ and  arbitrary $T$. We find that, in the $T'=\infty$ sector, the in-flux occurs only at the first spin.  In the $T<\infty$ sector (sites $x=1,2,...$), the out-flux shows a non-trivial profile: $F(x)$. Far from the junction of the two chains, $F(x)$ decays as $e^{-x/\xi}$, where $\xi$ is twice the correlation length of the {\em equilibrium} Ising chain. As $T\rightarrow 0$, this decay crosses over to a power law ($x^{-3}$) and resembles a ``critical'' system.  Simulations affirm our analytic results.}

\begin{document}

\title{Energy flux near the junction of two Ising chains at different temperatures}
\author{M. O. Lavrentovich\inst{1} \and R. K. P. Zia\inst{2}}
\maketitle

\section{Introduction}

Equilibrium
statistical mechanics, formulated by Boltzmann, Gibbs, and others about a century ago, 
is so well established that today, it is part of the
core material in typical physics programs. By contrast, statistical
mechanics of systems far from equilibrium is so poorly understood that the
Committee on CMMP 2010 of the National Research Council recently recognized
it as one of the six most fundamental and important challenges in condensed
matter and materials physics\cite{CMMP}. To be clear, we should distinguish
two types of such systems. On the one hand, we may encounter a statistical
system evolving with a dynamics which obeys detailed balance. Though it will
eventually settle into an equilibrium state and can be described within the
Boltzmann-Gibbs framework, its \emph{time-dependent} behavior can be highly
non-trivial. Good examples are slow relaxation or practically-stationary
states, such as aging \cite{aging} or glassy phenomena \cite{glass}. On a
more difficult level, we may face a system with a dynamics which violates
detailed balance. Examples abound in, e.g., the life sciences, sociology, or
economics, for which macroscopic variables are more appropriate. Typically,
the rules for their evolution do not obey time-reversal. Even when such a
system settles into a time-independent steady state, its stationary
distribution is unknown in general. Much of the intuition built on
equilibrium statistical mechanics fails under these circumstances. In an
effort to develop insight into this class of non-equilibrium steady states
(NESS), we take model systems with especially simple equilibrium properties
and drive them out of equilibrium by mechanisms which are physically
well motivated.

\begin{figure}
\onefigure[width=3.2in]{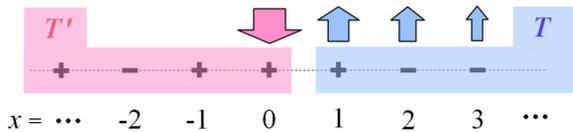}
\caption{We consider two semi-infinite chains of Ising spins (red and blue) that only assume the two values $\sigma_x = \pm 1$,  which we denote by ``$+$'' and ``$-$'' signs.  The two boxes labeled $T'$ and $T$, with $T' > T$, represent the two thermal reservoirs to which we couple the two Ising spin chains. The red (blue) arrows denote the energy flux flowing into (out of) the chain.  Narrower arrows denote a smaller flux. }
\label{f.setup}
\end{figure}

A good example of a simple model driven far from equilibrium is the kinetic
Ising model, coupled to two thermal reservoirs set at different temperatures 
\cite{Boring,others,RaczZia94,SS,LRW}. In previous studies, translational
invariance is kept as much as possible, e.g., every other spin being updated
according to the same reservoir \cite{others,RaczZia94}. As a result, the
effects of being driven out of equilibrium are extensive. In this article,
we study another, perhaps more common, way of coupling a system to two
baths, namely, one side being kept hot and the other, cold. Specifically, we
consider an Ising chain (i.e., a one-dimensional lattice with sites $x=...,-1,0,1,...$) 
in which all spins with $x\leq 0$ are coupled to one bath while the rest are coupled to another bath. Alternatively, we can view this system as joining together two Ising chains, each being in contact with its
own thermal bath. The coupling of the two spins at the ends ($x=0,1$) allows
the flow of energy from the hotter bath to the cooler one,  as illustrated in fig.~\ref{f.setup}.

Recently, a
simpler version of this model has been investigated \cite{Farago}. On a ring
of $2N+1$ spins (i.e., periodic chain), all but one is coupled to a standard
bath at \emph{zero temperature}, while one spin is flipped randomly, mainly
through a Poisson process. Energy is injected at the one spin, in that its
bonds are randomly broken. Meanwhile, these broken bonds, or domain walls,
wander into the rest of the ring and can annihilate each other when they
meet. Since the main interest is how the details of the Poisson flipping
control the fluctuations of injected energy, the focus turned to the
statistics of the location of the domain wall nearest the injection site.  
Thus, some information concerning the energy density profile is extracted.  
Since the domain wall statistics are the same as the statistics
of bonds being broken or not along the chain, 
the domain wall information can be extracted from the correlation of nearest neighbor spins. 
Our work here differs in mainly
two aspects. We find the entire two-spin correlation (not just the nearest
neighbor pairs) and indicate how that result provides us with correlations
for arbitrary numbers of spins. Secondly, our approach allows for the study of the
system set at any two temperatures, and we find explicit
results for the case where one part of the system is at infinite temperature and the other
is at an arbitrary temperature.

In the next section, the specifications of our model will be presented.
The system is no longer translationally invariant and the non-equilibrium
effects are localized. The third section will be devoted to theoretical analysis and comparisons with simulation results. By restricting ourselves to
``heat-bath'' dynamics\cite{G+HB}, we are able to solve for the correlation
functions analytically. With these, the energy flux  from
the hotter bath to the colder one can be investigated using the approach in 
\cite{RaczZia94}. Since such an effect is localized (around $x=0$), there is
a non-trivial ``profile'' of the average energy flow between the baths and
the chain. The details of this analysis are quite involved and will be
published elsewhere\cite{ML}. Here, we will only present the setup, some
key steps, and the results. In particular, for the most extreme case: $
T_{x<0}=\infty $ and $T_{x\geq 0}=0$, this profile vanishes for $x<0$ and
decays as $1/x^3$ for $x\gg 1$. We end with a summary and outlook for future
studies.

\section{The model}

In a kinetic Ising chain, spins are located on sites (labeled by an integer 
$x$) of a discrete lattice and assume only two values: $\sigma _x=\pm 1$.
Thus, a configuration of this system is described by the set of spins $%
\{\sigma _x\}$. Each spin interacts with its two neighbors via the
Hamiltonian: $\mathcal{H}\left( \{\sigma _x\}\right) =-J\sum_x\sigma
_x\sigma _{x+1}$, where $J$ is a coupling constant. For a finite system,
boundary conditions are needed, the simplest being periodic. 

While in contact with a thermal bath at temperature $T$, the kinetic Ising chain
has a simple
equilibrium distribution: $P_{\mathrm{eq}}\left( \{\sigma _x\}\right) =e^{-\mathcal{H}%
/k_BT}/Z$. Its static properties (e.g., the partition function $Z$) were
obtained first by Ising\cite{Ising}. Its dynamical properties, on the other
hand, are much more complex since they depend sensitively on how the system is
endowed with time dependence. The simplest is Glauber spin-flip
``heat-bath'' dynamics\cite{G+HB}. In a Monte Carlo simulation with parallel
updates, each spin ($\tilde{\sigma}_x$) is chosen and reset to be $\sigma _x$%
, with a probability that depends only on the average of its two neighboring
spins: $p_{x}\equiv \frac{1}{2}+\gamma \sigma_{x}\left( \tilde{\sigma}_{x-1}+%
\tilde{\sigma}_{x+1}\right) /4$, with $\gamma =\tanh (2J\slash k_BT)$. Thus,
with $J,\gamma >0$, the new spin will tend to be more aligned ($(
1+\gamma)/2$) with its neighbors than anti-aligned ($(1-\gamma)/2$).
The standard framework to deal with dynamics is based on the master equation
for a time-dependent $P\left( \{\sigma _x\},t\right) $. In this approach, we
have $P\left( \{\sigma_x\},t+1\right) =\sum_{\{\tilde{\sigma}
_x\}}\prod_{x} p_{x} P\left( \{\tilde{\sigma}_x\},t\right)$, and at large times,
the system settles into the stationary $P_{\mathrm{eq}}\left( \{\sigma
_x\}\right) $. A favorite alternative update is random sequential, where
only one randomly chosen spin is updated at each time step. We will use this
method for the ``two-temperature'' Ising chain below.

One way to drive this kinetic Ising chain out of equilibrium is to couple it
to more than one thermal bath. The simplest generalization is to have two
baths, at $T$ and $T^{\prime }>T$. Even with this simplification, there is
an infinite variety of ways to couple the spins to them. One is to update
each spin the same way, randomly choosing $T$ and $T^{\prime }$ with fixed
probabilities\cite{Boring}. On average, $\gamma $ remains homogeneous
and the intuitive picture of the system -- same as in equilibrium but with
an effective $\gamma $ -- proves to be correct. One may alternatively
 couple the two baths to every other spin\cite{RaczZia94,SS,MZS,LRW}, so
that we should write $\gamma_x$ in the $p_x$ above. The master equation
then takes the form 
\begin{equation}
P\left( \{\sigma _x\},t+1\right) =\sum_{\{\tilde{\sigma}_x\}}W\left(
\{\sigma _x\};\{\tilde{\sigma}_x\}\right) P\left( \{\tilde{\sigma}%
_x\},t\right) , \label{ME}
\end{equation}
where $W$ is given by
\begin{equation}
N_{{\rm tot}}^{-1}\sum_x\left[ \frac 12+\gamma _x\sigma _x\left( \frac{%
\tilde{\sigma}_{x-1}+\tilde{\sigma}_{x+1}}4\right) \right] \prod_{y\neq
x}\left[ \frac{1+\sigma _y\tilde{\sigma}_y}2\right].   \label{W}
\end{equation}
Here, $N_{\mathrm{tot}}$ is the total number of spins in the system, so that $N_{\mathrm{tot}}^{-1}\sum_x$ accounts for the possibility of any spin being chosen, with equal
probability, once per time step. The $\prod_{y\neq x}$ factor insures that
all other spins remain unchanged. For the model in \cite{RaczZia94}, we
have, say, $\gamma _{2n}=\tanh (2J\slash k_BT)$ and $\gamma _{2n+1}=\tanh (2J%
\slash k_BT^{\prime })$ with integer $n$. A prominent aspect of this model
is the non-trivial energy flux \emph{through} the system, flowing from the
hotter bath to the colder one. In particular, many of the properties of the
energy transfered to the baths due to the updating of the spins can be
computed exactly\cite{RaczZia94,LRW}. 
Entropy production (associated with the two
baths) can be defined and also computed. Due to the translationally invariant
nature of the alternating couplings, both the total flux and entropy
production are extensive.

In this study, we consider another way to couple the Ising chain to two
baths, namely, 
\begin{eqnarray}
\gamma _{x\leq 0} &=&\tanh (2J\slash k_BT^{\prime })\equiv \gamma ^{\prime }
\label{g'} \\
\gamma _{x\geq 1} &=&\tanh (2J\slash k_BT)\equiv \gamma  \label{g}
\end{eqnarray}
In other words, each of the two sectors of the chain is updated with a
single temperature, with the left sector being hotter ($T^{\prime }>T$).
This is a much more common form of coupling a system to two baths, occurring, for example, in
stovetop cooking. Of course, this model is still far from being realistic,
since heat is typically transported, via diffusion, from the hotter side to
the colder one. Here, due to the local nature of the spin-bath coupling,
non-trivial energy transfer is expected to take place close to the junction
and the anomalies associated with non-equilibrium statistical mechanics
should vanish for $\left| x\right| \gg 1$. Now, translational invariance is
broken and the analysis is more complex. In the next section, we provide a
solution to the problem, in terms of the two-spin correlation function $%
\left\langle \sigma _x\sigma _y\right\rangle $. Simulation data will also be
presented for the case $T^{\prime} = \infty$ ($\gamma' = 0$), 
showing a consistent picture of the inhomogeneous energy flux.

Before ending, let us remark that we will, for simplicity, ignore the
boundary conditions. A more rigorous way is to start with open boundaries,
write modified expressions for the end spins, and ensure that they make no
difference in the limit of large $N_{\mathrm{tot}}$. Alternatively, we can impose
periodic boundary conditions, deal with two junctions, and check that the
effects of the two disentangle appropriately.

\section{Theoretical analysis and simulation results}

For $W$'s which violate detailed balance, it is generally impossible to find
an explicit expression for the stationary distribution $P^{*}\left(
\{\sigma _x\}\right) $, i.e. the solution to 
$P^{*}=\sum WP^{*}$.  A formal solution exists\cite{Hill}, but it is so cumbersome that typically,
useful information cannot be extracted. An alternative approach is to seek
the correlations functions in the steady state
\begin{equation}
\left\langle \sigma _{x_1}\ldots\sigma _{x_k}\right\rangle \equiv
\sum_{\{\sigma _x\}}\sigma _{x_1}\ldots\sigma _{x_k}P^{*}\left( \{\sigma
_x\}\right).
\end{equation}
Inserting this definition into the equation for $P^*$, we face the sum 
$\sum_{\{\sigma _x\}}\sigma _{x_1}\ldots \sigma _{x_k}W\left(
\{\sigma _x\};\{\tilde{\sigma}_x\}\right)$ which, in general, involves $%
k^{\prime }>$ $k$ spins ($\tilde{\sigma}$'s). Writing an equation
for the $k^{\prime }$-spin correlation, we encounter more spins.  The result
is the BBGKY hierarchy \cite{BBGKY} involving correlations of arbitrarily
many spins. However, $W$ is linear in its arguments in our case
(a salient feature of ``heat-bath'' dynamics), so that coupled equations do not
proliferate.  
In particular, we find 
\begin{equation}
\sum_{\{ \sigma_x \}} \sigma _aW=\frac{\gamma _{a}}2\left( \tilde{\sigma}_{a-1}+\tilde{%
\sigma}_{a+1}\right) +\left( N_{{\rm tot}}-1\right) \tilde{\sigma}_a \,. \label{sum1}
\end{equation}
We now replace the dummy $\tilde{\sigma}_{a}$ by $\sigma _x$ and average both sides of eq.~\ref{sum1}.  The $N_{{\rm tot}}$ variable cancels and we obtain 
\begin{equation}
0=\gamma _x\left\langle \sigma _{x-1}+\sigma _{x+1}\right\rangle
-2\left\langle \sigma _x\right\rangle
\end{equation}
(with slight modification for the spins at the end of the chain).  Unless $\gamma_x = 1$, the only
solution to this equation is $\langle \sigma_x \rangle = 0$.  This result is hardly surprising, especially since an ordinary Ising chain displays no spontaneous magnetization for all positive
temperatures.

Turning to the two-point function, we first note 
\begin{equation}
\left\langle \sigma _x\sigma _x\right\rangle =1  \label{1}
\end{equation}
and 
\begin{equation}
\left\langle \sigma _x\sigma _y\right\rangle =\left\langle \sigma _y\sigma
_x\right\rangle  \label{xy}
\end{equation}
so that we can focus on, say, $x<y$ only. There, we obtain
\begin{align}
0&=\gamma _x\left( \left\langle \sigma _{x+1}\sigma _y\right\rangle
+\left\langle \sigma _{x-1}\sigma _y\right\rangle \right) \nonumber \\
& \quad +\gamma _y\left(
\left\langle \sigma _x\sigma _{y+1}\right\rangle +\left\langle \sigma
_x\sigma _{y-1}\right\rangle \right) -4\left\langle \sigma _x\sigma
_y\right\rangle \, . \label{2pf-eqn}
\end{align}
It is easy to verify that, for $\gamma _x=\gamma _y=\gamma $, a familiar
result emerges, i.e., $\left\langle \sigma _x\sigma _y\right\rangle
_{\mathrm{eq}}=\omega ^{\left| x-y\right| }$, with $\omega \equiv \tanh \left( J\slash k_BT\right)$.

Before we solve equation eq.~(\ref{2pf-eqn}), let us remark that all correlations with an odd
number of spins vanish. Further, with our dynamics, correlations of \emph{%
any even} number of spins can be expressed in terms of products of $%
\left\langle \sigma _x\sigma _y\right\rangle $ \cite{RaczZia94,SS,MZS,A,MM}.
In this sense, we have a complete picture of this non-equilibrium system.

\begin{figure}
\onefigure[width=2.0in]{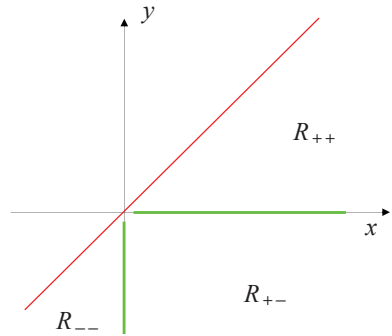}
\caption{The domains $R_{--}$, $R_{++}$, and $R_{+-}$ in which we solve eq.~\ref{2pf-eqn} for $\left\langle
\sigma _x\sigma _y\right\rangle$.  We have the BC $\left\langle
\sigma _x\sigma _x\right\rangle =1$ along the red line and match the solutions along the green lines.}
\label{f.domains}
\end{figure}
Returning to our problem, we see it is a generalized, discrete Helmholtz equation and so, boundary conditions (BCs) are crucial. One BC is eq.~\ref{1}  and the other
is $\left\langle \sigma _x\sigma _y\right\rangle \rightarrow 0$ for $\left|
x-y\right| \rightarrow \infty $. In other words, we have Dirichlet BC's for
the half space $x\geq y$. In our case, we have only two $\gamma $'s. Thus,
this domain can be partitioned into three regions, $R_{++}$, $R_{--}$, and $%
R_{+-}$ (corresponding to $x,y$ being positive or negative, as shown in fig.~\ref{f.domains}), 
in each of which the operator is \emph{homogeneous}. Thus, the solution
can be attacked by standard, though tedious, methods. We will only indicate
some key points for the general problem and provide a few more details below
for the special case of $T^{\prime }=\infty $, i.e., $\gamma ^{\prime }=0$.

Note that the operator in eq.~(\ref{2pf-eqn}) is isotropic in $R_{++}$ and $%
R_{--}$, but anisotropic in $R_{+-}$. Since the operators are homogeneous
(discrete) Laplacians within each region, we can make use of the
eigenfunctions $e^{ikx+ipy}$ with eigenvalues $2\left( \gamma _x\cos
k-1\right) +2\left( \gamma _y\cos p-1\right)$.
In the regions $R_{--}$ and $R_{++}$, appropriate combinations must be chosen to satisfy
eqs.~(\ref{1}, \ref{xy}). The last BC is less straightforward and we will
rely on the following approximation. We choose functions which vanish on the $%
x=N$ line in $R_{++}$ and the $y=-N$ line in $R_{--}$ first, and then let $%
N\rightarrow \infty $ at the end. Such an approach can also be exploited in $%
R_{+-}$. Finally, we must match the functions on the common boundaries ($x=0$
and $y=0$ lines, i.e. the green lines in fig.~\ref{f.domains}).

Let us consider an easier case, $\gamma ^{\prime }=0$, so we have
only one parameter left: $\gamma$. Eq. (\ref{2pf-eqn}) immediately
provides 
\begin{equation}
\left\langle \sigma _x\sigma _y\right\rangle =0\quad \mbox{for}\quad y<x\leq 0
\label{xy0}
\end{equation}
which is quite reasonable for $R_{--}$ where spins are flipped completely
randomly. Next, in $R_{+-}$, this equation
simplifies to a single one, $\left\langle \sigma _x\sigma
_{y+1}\right\rangle +\left\langle \sigma _x\sigma _{y-1}\right\rangle
=4\left\langle \sigma _x\sigma _y\right\rangle /\gamma $, for \emph{all }$%
y\leq 0$ (and $x>y$). It is satisfied by $\left\langle \sigma _x\sigma
_y\right\rangle =e^{\mu x}$, with (real) $\mu =\pm \cosh ^{-1}\left(
2/\gamma \right) $. To be careful, let us first focus on $y<0$. From eq.~(\ref{xy0}) 
above, we have $\left\langle \sigma _0\sigma _{y<0}\right\rangle
=0$, which restricts us to $\left\langle \sigma _x\sigma _y\right\rangle
=A\sinh \mu x$. Imposing the BC for $x\rightarrow \infty $, we arrive at 
\begin{equation}
\left\langle \sigma _x\sigma _{y<0}\right\rangle =0\quad \forall \,\,x>y \, .
\label{x<0}
\end{equation}
On the line $y=0$, we have a different BC, since $\left\langle \sigma
_0\sigma _0\right\rangle =1$. The other BC then picks out 
\begin{equation}
\left\langle \sigma _x\sigma _0\right\rangle =\tilde{\omega}^x\quad \forall
\,\,x\geq 0  \label{x=0}
\end{equation}
where 
\begin{equation}
\tilde{\omega}\equiv \frac 2\gamma -\sqrt{\left( \frac 2\gamma \right) ^2-1} \,.
\label{omega-tilde}
\end{equation}
In other words, this particular correlation is of the same form as that for
the Ising chain in equilibrium, with an effective temperature given by $%
\gamma /2$ instead of $\gamma $. Intuitively, this result is appealing,
since one spin ($\sigma _0$) is coupled to $\gamma ^{\prime }=0$ and the
other ($\sigma _y$) is coupled to $\gamma $.

Finally, we turn our attention to $R_{++}$. Here, eq.~(\ref{2pf-eqn}) is
exactly the same as for the equilibrium Ising chain. Indeed, the \emph{only}
difference between the two problems is the BC on the $y=0$ line, where we
have eq.~(\ref{x=0}) instead of $\left\langle \sigma_x \sigma_0\right\rangle
_{\mathrm{eq}}=\omega ^x$. Thus, we can simplify our problem by considering the
difference 
\begin{equation}
S\left( x,y\right) \equiv \left\langle \sigma _x\sigma _y\right\rangle
-\left\langle \sigma _x\sigma _y\right\rangle _{\mathrm{eq}} 
\end{equation}
which \emph{vanishes} on two of the three boundaries of $R_{++}$. A further
advantage is $S$ is directly related to the quantity of interest -- the
flux, $F\left( x\right) $, i.e., the average rate of energy loss when $%
\sigma _{x>0}$ is updated. Specifically, from \cite{RaczZia94}, one can show that
\begin{equation}
F(x)= \frac{1}{4} \left[\gamma S\left( x+1,x-1\right) -S\left( x+1,x\right) -S\left(
x,x-1\right)\right]
\end{equation}
in units of $4J$.  In other words, once we find $S$, the energy flux profile can be computed.

Our problem now reduces to solving the Dirichlet problem
\begin{equation}
\mathcal{D}S=0\quad ;\quad \mathcal{D}\equiv 4\left( \frac 1\gamma -1\right)
-\Delta _{x,y}
\end{equation}
in $\,R_{++}$ with BC's 
\begin{equation}
S\left( x,0\right) =\tilde{\omega}^x-\omega ^x\quad ;\quad S\left(
x,x\right) =S\left( N,y\right) =0
\end{equation}
Here,  $\Delta _{x,y}$ is the discrete Laplacian and $N$ will be taken to $%
\infty $ at the end. A standard route is to use the Dirichlet Green's
function $G$ which satisfies $G=0$ on the boundaries and 
$\mathcal{D}G(x,y;\xi ,\eta )=\delta _{x,\xi }\delta _{y,\eta }.$ The result is 
\begin{align}
G(x,y;\xi ,\eta )& =\frac{2\gamma }{N^2}\sum_{m=1}^{N-1}\sum_{n=1}^{m-1}
\frac{U_{k,p}\left( x,y\right) U_{k,p}\left( \xi ,\eta \right) }{2-\gamma (\cos k+\cos p)} \,,  \label{green}
\end{align}
where the eigenfunctions of our operator ($\mathcal{D}$ and BC's) are 
\begin{equation}
U_{k,p}\left( \xi ,\eta \right) =\sin k\,\xi \,\sin p\,\eta -\sin p\,\xi
\,\sin k\,\eta 
\end{equation}
with $\left( k,p\right) \equiv \pi \left( m,n\right) /N$. Exploiting the identity
$S = \sum [S \mathcal{D} G - G \mathcal{D} S]$ and a discrete
divergence theorem, we find 
\begin{align}
S(x,y) =\sum_{\xi =1}^{N-1}\left[ G(x,y;\xi ,1)\left( \tilde{\omega}^\xi -\omega
^\xi \right) \right].
\end{align}
Applying $4/\gamma = \tilde{\omega} + \tilde{\omega}^{-1}$ and 
$2/\gamma = \omega + \omega^{-1}$, we arrive at an explicit solution in $
R_{++}:$ 
\begin{align}
&S(x,y) =\frac{\gamma ^3}{N^2}\,\sum_{m,n}U_{k,p}\left( x,y\right) \left[ 
\frac{\sin k\sin p\,\left( \cos k-\cos p\right) }{2-\gamma (\cos k+\cos p)}%
\right]  \nonumber \\
&\times  \frac{\left[\gamma (\cos k+\cos p)-3\right]}{\left( 2-\gamma \cos
k\right) \left( 2-\gamma \cos p\right) \left( 1-\gamma \cos k\right) \left(
1-\gamma \cos p\right) }
\end{align}
In the $N\rightarrow \infty $ limit, $p,k$ become continuous in $[0,\pi ]$
and convenient variables are $\theta ~\equiv \left( k+p\right) /2$ and $\phi
\equiv \left( k-p\right) /2$. Inserting the result into the expression for $F
$, the flux profile can be written as a double integral 
\begin{equation}
F(x)=\frac{\gamma ^3}{8\pi ^2}\int_0^\pi \,\int_0^{\pi -\phi }\mathrm{d}
\theta \, \mathrm{d}\phi \,A\left(\theta,\phi\right) \sin 2\theta x \, ,
 \label{F1}
\end{equation}
where $x=1,2,\ldots$ and
\begin{align}
& A\left( \theta ,\phi \right) \equiv \frac{\sin \theta \sin ^2\phi \left[
\sin ^2\phi -\sin ^2\theta \right] \left[ \gamma \cos \phi -\cos \theta
\right] }{1-\gamma \cos \theta \cos \phi } \nonumber \\
&\times \sum_{a=1}^2\frac{(-1)^a}{2a^2-4\gamma a\cos \theta \cos \phi
+\gamma ^2(\cos 2\theta +\cos 2\phi )}
\end{align}

We now consider the large $x$ asymptotics of $F$. Deferring a detailed presentation
to elsewhere \cite{ML}, we provide only some results here.  We find that eq.~\ref{F1} 
for integers $x \geq 2$ reduces exactly to
\begin{equation}
F(x)  = \frac{1}{\pi \gamma^{2x+1}} \int_0^{1}\frac{\left(1- \gamma ^2 \eta ^4\right)  \left(1 - \sqrt{1-\gamma^2 \eta^2}\right)^{2x}}{\eta^{2x}\left(1 -\gamma^2 \eta ^2+  \gamma ^2 \eta ^4 \right) \sqrt{1- \eta^2}} \, \mathrm{d} \eta ,\label{fintegral}
\end{equation}
with a known correction for $x = 1$ \cite{ML}.  
To get the cross-over behavior as $\gamma \rightarrow 1$ ($T \rightarrow 0$), 
we can arrange an asymptotic expansion of eq.~\ref{fintegral} (for large $x$)
as a sum over modified Bessel functions 
\begin{align}
F(x)=\omega ^{2x}\sum_{n=0}^\infty B_n\,\zeta ^{-n}e^\zeta K_{n+1}\left(
\zeta \right),  \label{F-K}
\end{align}
where $\zeta \equiv 2\sqrt{1-\gamma ^2}x $ and $B_n$ are
explicitly known coefficients.

Intuitively, we expect $F$ to decay exponentially for $T>0$, where the
equilibrium correlation length, $\xi \equiv -1/\ln \omega $, is finite.
Careful analysis \cite{ML}
confirms this picture. In particular, eq.~(\ref{F-K}) leads to
\begin{equation}
F(x)\rightarrow \, \frac{1}{4}\sqrt{\frac{(1-\omega ^2)^5}{\pi \omega^2 (1+\omega ^2)^3}} \,\left[\frac{\omega ^{2x}}{\sqrt{x}} \right]\left\{ 1+O\left( x^{-1}\right)
\right\} \,.   \label{asymptotic}
\end{equation}
Interestingly, we observe that the decay length is $\xi/2$, 
though we have no good heuristic explanation. 
In the extreme case of $T=0$, these complex expressions simplify and
a power law appears.  The result is
\begin{equation}
\lim_{\gamma \rightarrow 1}F(x)=\frac {1}{2\pi x^3}.  \label{powerlaw}
\end{equation}
Though reminiscent of the crossover to critical exponents, we also have no simple
argument in favor of the specific value $-3$. Of course, the total steady flux is
finite, so that $\Sigma _{x>0}F\left( x\right) $ must converge. But this
constraint only limits the power to be less than $-2$. 

\begin{figure}
\onefigure[width=2.8in]{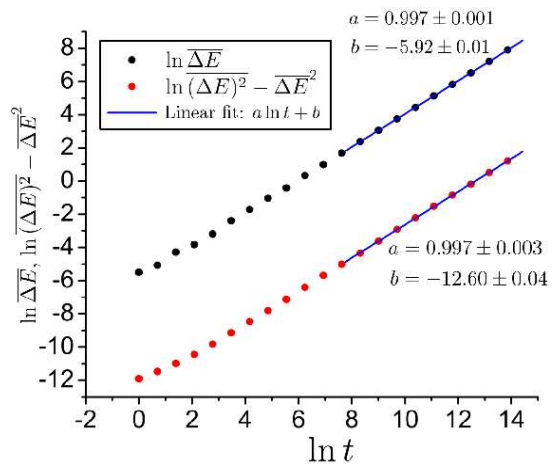} 
\caption{The average (over $10^4$ independent simulation runs) of the change in energy $\overline{\Delta E}$  and the variance $\overline{(\Delta E)^2}-\overline{\Delta E}^2$ at spin $x = 3$ as a function of the number of times $t$ that the spin was updated per run. Here, $\gamma = 0.9$ (with $T' = \infty$).}
\label{f.randomwalk}
\end{figure}
\begin{figure}
\onefigure[width=2.7in]{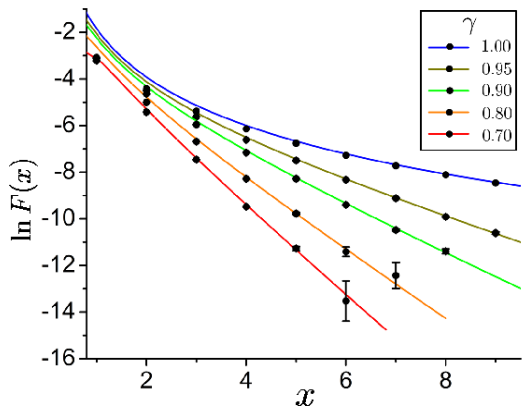} 
\caption{The dots are the energy flux $F(x)$ at spin $x$ computed via Monte Carlo simulations (with error bars smaller than the dots for $\gamma \geq 0.9$).  The solid lines are the first two orders of the expansion in eq.~\ref{F-K}.}
\label{f.fluxes}
\end{figure}

As further confirmation of these analytic results, we performed Monte-Carlo
simulations of our system. For $T^{\prime }=\infty $, it is unnecessary to
include any spin at $x<0$. Instead, we study a chain with $N+1$ spins, the
first ($\sigma_0$) being coupled to the infinite temperature bath. At each
step, a random spin is chosen and assigned a new value according to the
probabilities given above, except $\sigma _0$, which is assigned $\pm 1$
randomly. Using $N\approx 30$, we evolved the system for approximately $
10^{6}$ time steps (i.e. updates per spin). 
When a spin flips during an update, we record the
change in the energy $\Delta E$, which can take only the values $0$ and $\pm
1$ (in units of $4J$). In this manner, we have a time series $\Delta E\left( x,t\right) $ for
each site. It is clear that, over a run, this quantity essentially performs
a random walk. For systems in equilibrium, such walks will be unbiased. For
our case, we expect \emph{biased} walks. To verify this scenario, we carried
out $10^4$ independent runs and computed both the average, $\overline{\Delta
E\left( x,t\right) }$, and the variance $\overline{\left[ \Delta E\left(
x,t\right) \right] ^2}-$ $\left[ \overline{\Delta E\left( x,t\right) }%
\right] ^2$.  As illustrated in fig.~\ref{f.randomwalk}, both the average and the variance
increase with $t$ linearly (i.e., $a\cong 1$). The intercept, $b$, of the former provides
us with data to compare with $F(x)$. Similarly, the intercept of the latter
allows us to estimate the error bars in fig.~\ref{f.fluxes}.
 Fig.~\ref{f.fluxes} shows the excellent agreement with
theoretical predictions and simulations. 

Finally, we remark that we did not present results for the \emph{in-flux} of
energy, which occurs only at $x=0$, since there are no correlations
involving $\sigma_{x<0}$. It is clear that its value must be just $\Sigma
_{x>0}F\left( x\right) $.  

\section{Summary and Outlook}

In this article, we consider joining together two Ising chains end-to-end,
each coupled to a uniform (but different) temperature reservoir. As a
result, when the system settles into a steady state, there is a constant flow
of energy through the system, from the hotter to the colder bath. Far from
the junction, each sector of the chain is expected to behave as one in
equilibrium, so that the effects of this energy flow are localized. 
Such a through-flux is strictly a non-equilibrium phenomenon and the
stationary distribution is not simply given by some spatially dependent,
effective Boltzmann factor. By choosing heat-bath spin-flip dynamics, we
are able to compute all the correlation functions. In particular, we provide
key points for the calculation and the results for the two-spin correlation, $%
\left\langle \sigma _x\sigma _y\right\rangle $. This is not simply a
function of $\left| x-y\right| $, since our system is not translationally
invariant. From these functions, we can compute $F\left( x\right) $, the
steady flux of energy out of the system, as a result of updating the spin at 
$x$. When one reservoir is set at $T^{\prime }=\infty $, we find explicitly
that $F\left( x\right)$ decays into the sector with $T<\infty $
exponentially, with a length half of the correlation length of its
equilibrium counterpart. As $T\rightarrow 0$, there is no such counterpart,
as the decay crosses over to a power law: $1/x^3$. Simulations of this model
verify these results.

Before closing, let us consider possible avenues for future studies, as well
as related Ising-like systems far from equilibrium. The most immediate case
of interest is finite $T^{\prime }$, analytic results for which should be
within reach. There, we expect exponentials for $F\left( x\right)$ in both
sectors. It is not clear if the decay lengths continue to be as simple as
what we found. In particular, $F$ must vanish as $T^{\prime }\rightarrow T$.
Whether this limit appears as an overall amplitude which vanishes,
analytically with $\left( T^{\prime }-T\right) ^2$ or more singularly, would
be interesting. Beyond the Ising chain, the most obvious generalization is
the same model in higher dimensions, joining two systems side by side, say.
Since the ordinary Ising model at equilibrium displays phase transitions in higher dimensions, more interesting
questions arise. Examples include: Does $F$ decay with singular powers when $T$ is set at $T_c$, the critical temperature (with $T^{\prime }\gg T_c$)?
If so, are there new exponents (such as the dynamic exponent $z$ in model A
\cite{HH})? or new combinations of existing exponents (such as $z=4-\eta $
in model B\cite{HH})? Setting $T<T_c$, we could expect a non-trivial
magnetization profile $\left\langle \sigma _x\right\rangle $. How does it
decay with $x$? Continuing along these lines, are there new singularities
associated with this profile when $T^{\prime}$ is lowered to $T_c$?

Another generalization of joining two Ising systems is also immediate: using
Kawasaki spin-exchange dynamics\cite{Kawa} instead. While no new static
properties are present in an equilibrium Ising model with this type of
dynamics, much more surprising behavior for the two-temperature case have
been observed, for both 1- and 2-d systems \cite{MSZ}. Beyond these, there
are limitless varieties, most of which will undoubtedly provide further
insight into how we may proceed in attempting to formulate an overarching
principle for non-equilibrium statistical mechanics.

\acknowledgments

We thank B. Derrida, M. Pleimling, and B. Schmittmann for illuminating discussions. 
One of us (RKPZ) is grateful for the hospitality of H. W. Diehl in the Universit\"{a}t
Duisburg-Essen, where some of this work is done. This research is supported
in part by the Alexander von Humboldt Foundation and by USNSF, DMR-0705152.

\end{document}